\documentstyle[seceq,preprint,epsf]{ptptex}

\def\Lgr{{\cal L}}
\newcommand{\vev}[1]{\langle#1\rangle}
\def\be{\begin{equation}}
\newcommand{\bel}[1]{\begin{equation}\label{#1}}
\def\ee{\end{equation}}
\newcommand{\eref}[1]{(\ref{#1})}
\newcommand{\Eref}[1]{Eq.~(\ref{#1})}
\newcommand{\rem}[1]{}

\def\half{{1\over 2}}
\def\MM{{\cal M}}
\def\NN{{\cal N}}

\def\none{$\NN=1$}
\def\ntwo{$\NN=2$}

\def\nfour{$\NN=4$}
\def\susy{supersymmetry}
\def\susic{supersymmetric}

\def\rarr{\rightarrow}

\def\ZZ{{\bf Z}}
\def\ZN{\ZZ_N}

\def\NO{Nielsen-Olesen}
\def\ures{{\underline {\it Result}:} }
\preprintnumber{
IASSNS--HEP--98/24 \\ hep-th/yyymmnn}

\title{
Messages for QCD from the Superworld\footnote{Talk given at YKIS'97, Kyoto, Japan.}
}

\author{
Matthew J.~{\sc Strassler}\footnote{E-mail address: strassler@ias.edu}
}

\inst{
School of Natural Sciences, Institute for Advanced Study\\
Olden Lane, Princeton, NJ 08544  USA}

\recdate{
\today
}

\abst{
Recent discoveries in supersymmetric gauge theories have significant
implications for our understanding for QCD and of field theory in
general.  The phases of \none\ supersymmetric QCD (SQCD) are
discussed, and the possibility of similar phases in non-supersymmetric
QCD is emphasized.  It is described how duality in SQCD links many
previously known duality transformations that were thought to be
distinct, including Olive-Montonen duality of \nfour\ supersymmetric
gauge theory and quark-hadron duality in (S)QCD.  A link between
Olive-Montonen duality and the confining strings of (S)QCD is
explained, in which a picture of confinement via non-abelian monopole
condensation --- a generalized dual Meissner effect --- emerges explicitly.
In this picture, unlike previous ones, the confining flux tubes carry the
correct $\ZN$ discrete charges.  A number of studies of these subjects,
which could be carried out using lattice gauge theory, are proposed.
}

\begin{document}

\maketitle

\section{Introduction}

In this talk, I would like to describe some of the important lessons
that \none\ supersymmetric field theories teach us about gauge
theories in general.  While direct applications to QCD are few, there
are nonetheless important qualitative insights which can be gained.
In this talk I will mention three.  The first involves our new
understanding of the complexity of the phase structure of \none\
\susic\ QCD (SQCD), which raises the question, ``what is the phase
structure of QCD, and are there lattice or analytic approaches that
will efficiently allow us to find out?''  A second topic involves the
conceptual unification of many superficially different duality
relations, such as Olive-Montonen duality of \nfour\ \susic\ gauge
theories, the dual Meissner effect as a model of confinement, and the
relation between a theory of quarks and gluons and its description
using a chiral Lagrangian.  Although these naively seem unrelated,
they and others are all special cases of \none\ duality.  I will then
present some new results linking the QCD string and the Olive-Montonen
duality of \nfour\ \susic\ QCD.  I will show how solitonic QCD strings
arise in this context {\it without} any abelian projection, in the
context of a generalized, non-abelian dual Meissner effect.  I will
discuss some possible problems with approaches based on the abelian
projection. I'll also have a few things to say about QCD string
tensions, as calculated in field theory and in M theory, and will
emphasize that lattice QCD theorists should compute ratios of string
tensions in $SU(N)$ for $N\geq 4$.  As an aside, I'll mention some related
work in which the first example of the non-abelian dual Meissner
effect was presented.  Finally, I'll describe what I think are
the key questions raised by the superworld for QCD theorists.

\section{Phases of SQCD}

Following the work of Seiberg \cite{NAD} and
others,\cite{kinstwo,kinsrev,kutsch,ils,gtwo} we now know that
SQCD has many phases.  For my purposes I will limit myself to five
which both are common\cite{NAD} and are likely to be seen in
non-\susic\ theories.  (I will ignore the Higgs phase,
which is of course well-understood.)  These are the following:

1) The Free Electric Phase: the theory has so much matter that its
beta function is positive and it is free in the infrared.

2) The Non-Abelian Coulomb Phase: the theory is asymptotically free,
but its beta function hits a zero at a finite value of the gauge
coupling.  The low-energy theory is an infrared fixed point, an
interacting conformal field theory, which has no particle states.  Its
operators have non-trivial anomalous dimensions, some of which can be
exactly computed.

3) The Free Magnetic Phase: the original theory is asymptotically
free, and its coupling grows monotonically. The low-energy physics can
be described in terms of composite fields, of spin $0$, $\half$, and
$1$, many of which are non-polynomial --- indeed, non-local --- in terms
of the original fields.  The theory describing these composites is
again a gauge theory, with its own conceptually separate gauge symmetry!  
The dual theory has a positive beta function, so it is free in the
infrared and is a good perturbative description of the low-energy
physics.

4) Confinement Without Chiral Symmetry Breaking: in this
case the low energy theory is an infrared-free {\it linear} sigma
model, whose composites, of spin $0$ and $\half$, are
polynomial in the original fields.  

5) Confinement With Chiral Symmetry Breaking: similar
to the previous, except that the low-energy theory is
a {\it non-linear} sigma model, as in QCD.

Some comments: phases (1) and (5) are well-known.  Phase (4) has been long
suspected and debated as a possibility in some gauge theories.  Phase
(2) can be seen in perturbation theory, both in QCD and SQCD, when the
number of colors and flavors is large and the one-loop beta function
is extremely small by comparison; such fixed points are often called
Banks-Zaks fixed points,\cite{bz} though they were discussed by
earlier authors as well.  What is new here is that this phase exists
far beyond perturbation theory into an unexpectedly wide range of
theories, as we will see in a moment.  Phase (3) is entirely new and
previously unsuspected; it is perhaps the most spectacular of the
recent results.

It is also worth noting that while calculational techniques exist for
studying the infrared physics in most of these phases, the non-abelian
coulomb phase requires an understanding of four-dimensional
superconformal field theory.  Techniques in this subject are still
being explored\cite{CFT} and there is much left to be learned.

As an example, consider $SU(N_c)$ SQCD with $N_f$ flavors in the
fundamental and anti-fundamental representations.  This theory is in
the same universality class as (and is said to be ``dual'' to)
$SU(N_f-N_c)$ with $N_f$ flavors and with $N_f^2$ gauge singlets which
interact with the flavors.\cite{NAD} This dual theory serves as the
low-energy gauge theory in the free magnetic phase.  The five phases
appear for the following ranges of $N_c$: phase (1), $N_f\geq 3N_c$;
phase (2), $3N_c>N_f>{3\over2}N_c$; phase (3), ${3\over2}N_c\geq N_f
>N_c+1$; phase (4), $N_f=N_c+1$; phase (5), $N_f=N_c,0$.  For
$N_c>N_f>0$ the theory has no stable vacuum.  (Note that
supersymmetric theories generally have continuous sets of inequivalent
vacua; for each theory I have only listed the phase of the vacuum with
the largest unbroken global symmetry.)

As another example, consider $SO(N_c)$ with $N_f$ flavors in the
vector representation, whose dual is
$SO(N_f-N_c+4)$.\cite{NAD,kinstwo}  In this case we have phase (1),
$N_f\geq 3(N_c-2)$; phase (2), $3(N_c-2)>N_f>{3\over2}(N_c-2)$; phase
(3), ${3\over2}(N_c-2)\geq N_f >N_c-3$; phase (4), $N_f=N_c-3, N_c-4$.
As before $N_f<N_c-4$ has no vacuum, except $N_f=0$ which has vacua
with both phase (4) and (5).

Note that the word ``confinement'' has been used loosely here.  The
cases $SU(N_c)$ with $N_f=N_c+1,N_c$ are examples of
``complementarity'', where the confining and Higgs phases are actually
two regions in a single phase.\cite{compl}  There is no Wilson loop
with an area law; all sources can be screened by the massless fields,
and so no confining string can form.  By contrast, a spinor-valued
Wilson loop detects the confinement in $SO$ theories with vectors,
while Wilson loops in, for example, the ${\bf N_c}$ representation,
can detect the confinement in the pure $SU(N_c)$ and $SO(N_c)$ SQCD.

Aside from these two examples, many others are known, with
qualitatively similar phase diagrams.  Various new phenomena
have been uncovered.  But most \none\ \susic\
field theories are not understood, and much work remains to be done.

The most remarkable aspect of these phase diagrams is that they show
that the phase of a theory depends on (a) its gauge group $G$, (b) its
massless matter representations $R$, and although not shown here, (c)
its interactions $\Lgr_{int}$, renormalizable and non-renormalizable.
The dependence on $R$ goes far beyond the mere contribution of the
matter to the beta function; the matter fields are clearly more than
spectators to the gauge dynamics.  (A quenched approximation could not
reproduce this phase structure.)  The dependence on non-renormalizable
interactions is familiar from technicolor theories: a higher-dimension
operator, though irrelevant in the ultraviolet, may become relevant in
the infrared and control the physics of the low-energy theory.

Given this is true for SQCD, why should it not be true for non-\susic\
gauge theories?   Phases (1), (2) and (5) certainly arise.  It would
be remarkable indeed if the ubiquity of phase (4) and of the existence
of phase (3) could be demonstrated.  There might also be as yet
unknown phases that do not occur in \susic\ theories. 

 More specifically, we should seek to answer the following question:
what is the phase of QCD as a function of $G$, $R$ and $\Lgr_{int}$?
Unfortunately the answer cannot be learned from the supersymmetric
theories: the process of breaking supersymmetry leads to ambiguities
in the duality transformations.  We therefore need new tools, both
analytical and numerical.  This is clearly an area for lattice gauge
theory, but it is not easy to study the renormalization group flow
over large regions of energy using the lattice.  Additional analytic
work is needed to make this more tractable.  I hope some readers
will be motivated to consider this problem!

 It should be stressed that this is not merely an academic question.
It is possible that the correct theory of electroweak supersymmetry
breaking (or of fermion masses, etc.) has not yet been written down.
Perhaps a modified form of technicolor or something even more exotic
will appear in the detectors of the Large Hadron Collider, in a form
that we will be unable to understand unless the questions raised
above are addressed in the coming years.

\section{Unification of Dualities in \none\  Supersymmetry}

Let me begin by listing some duality transformations.

\

Electric-Magnetic (EM): this is the usual duality transformation of
the Maxwell equations without matter, which can be trivially extended
to \nfour, $2,1$ SQED.  The electric and magnetic gauge groups are
$U(1)_e$ and $U(1)_m$ (note these two symmetry groups are {\it
completely distinct} transformations on the non-locally related
electric and magnetic gauge potentials.)  The electric and magnetic
couplings are $e$ and $4\pi/e$.  (The last relation is modified for
non-zero theta angle.)

Dual Meissner (DM): for abelian gauge theory, or for a non-abelian
gauge theory which breaks to an abelian subgroup.  The theory has
magnetic monopoles, which are described by a magnetic abelian gauge
theory as ordinary charged particles.  The monopoles condense,
breaking the magnetic gauge symmetry, screening magnetic flux and
confining the electric flux of the original theory.

Olive-Montonen\cite{omdual,gom,osborn} (OM): the EM case for \nfour\
\susy, extended to a non-abelian gauge group $G_e$.  The magnetic
variables also are an \nfour\ \susic\ gauge theory and have a gauge
group $G_m$.  The theory is conformal and has a non-running coupling
constant $g$ in the electric theory and $4\pi/g$ in the magnetic
theory.  (The last relation is modified for non-zero theta angle.)

Generalized Dual Meissner\cite{rdew,spinmono} (GDM): similar to the DM
case, but where both the electric and magnetic gauge groups $G_e$ and
$G_m$ are non-abelian.  Condensing magnetically charged monopoles
again break $G_m$, screen magnetic flux and confine electric flux.

Seiberg-Witten pure\cite{nsewone} (SWp): for pure \ntwo\ \susic\
Yang-Mills theory.  The electric theory has gauge group $G$ of rank
$r$, whose maximal abelian subgroup is $[U(1)^r]_e$.  EM duality
applies to each $U(1)$; the magnetic theory has gauge group
$[U(1)^r]_m$.

Seiberg-Witten finite\cite{nsewtwo} (SWf): for a finite \ntwo\ \susic\
gauge theory with matter.  Very similar to OM above, but in general
the relation between $G_e$ and $G_m$ differs from the OM case.

QCD and the Sigma Model (QCD$\sigma$): here a strongly-coupled,
confining QCD or \none\ SQCD theory is described in terms of gauge
singlets, using a linear or non-linear sigma model.  This is not
always considered a ``duality'', but as we will see, it should be.

The main point of this section is to emphasize that all of these
dualities are linked together\cite{NAD,kinsrev} by results in \none\
\susy.  This can be easily seen using the duality of \none\ $SO(N)$
gauge theories with $N_f$ fields in the vector representation; as
mentioned in Sec. 2, such theories are dual to $SO(N_f-N+4)$ with
$N_f$ vectors and $N_f(N_f+1)/2$ gauge singlets.

Consider first $SO(2)$ without matter ($N_f=0$).  The dual theory is
again $SO(2)$ without matter --- EM duality --- which justifies
referring to the dual theory as ``magnetic''.

Next, consider $SO(3)$ with one triplet; its magnetic dual is $SO(2)$
with fields of charge $1,0,-1$ coupled together.  These theories are
both \ntwo\ \susic; the electric theory is the pure \ntwo\ $SU(2)$
theory studied by Seiberg and Witten, and the magnetic dual is the
theory of the light monopole which serves as its low-energy
description.\cite{nsewone} Since a mass for the triplet leads to
confinement of $SU(2)$ via abelian monopole
condensation,\cite{nsewone} this example gives both SWp and DM
duality.

Finally, take $SO(3)$ with three triplets, whose dual is $SO(4)\approx
SU(2)\times SU(2)$, with three fields in the ${\bf 4}$ representation
and six singlets.  If the triplets are massive, the quartets of the
dual theory condense, $SO(4)$ is broken, and confinement of $SO(3)$
occurs\cite{NAD} --- GDM duality.  The low-energy description below
the confinement scale is given by this broken $SO(4)$
theory\cite{NAD}, a non-linear sigma model --- QCD$\sigma$ duality.
And if instead the triplets are massless and are given the
renormalizable interactions which make the theory \nfour\ \susic, the
dual theory is consistent with OM duality:\cite{omdual,nsewtwo} one of
the $SU(2)$ subgroups of $SO(4)$ confines, leaving a single $SU(2)$
factor with three triplets coupled by the required \nfour\ \susic\
interactions.\cite{kinstwo,kinsrev}

\begin{figure}
\centering
\epsfxsize=3in
\hspace*{0in}
\epsffile{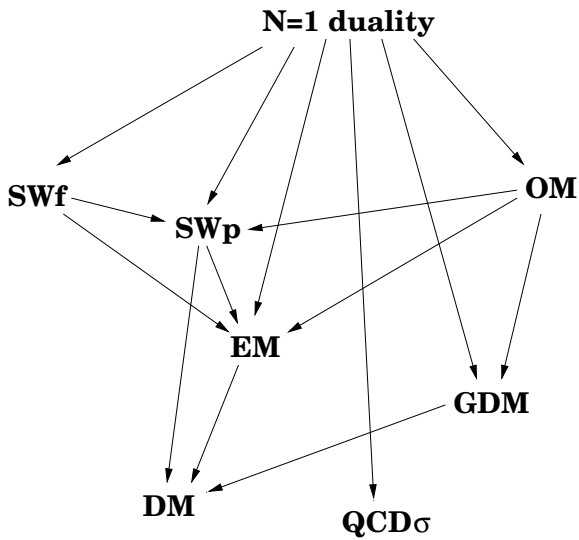}
\caption{Interrelation between types of duality.}
\label{fig:dualities}
\end{figure}

Thus, \none\ duality links all of the four-dimensional dualities on
the list together, showing they are manifestations of a single
phenomenon.  Figure \ref{fig:dualities} shows a cartoon of the
connections between these theories.  I have omitted, both from my list
and from this figure, other types of field theory duality that were
not known until recently.  Also not shown are the relations between
these dualities and those of the-theory-formerly-known-as-string
theory, sometimes termed M theory.  The situation brings to mind the
old story of the blind men and the elephant, in which each man feels
one part of the elephant --- the ear, or the trunk, or the tail ---
and erroneously assigns it a special significance, not realizing that
the apparently different parts are connected to a larger whole.
Having discovered the existence of a larger whole, we find ourselves
compelled to explain duality in a unified way.

I want to emphasize to those who are not familiar with this subject
that this picture, while not proven, is by no means speculative.  The
circumstantial evidence in its favor --- a vast number of consistency
conditions --- is completely overwhelming; one could easily give
twenty lectures on this subject.  A proof is still badly needed,
however; the underlying meaning of duality remains mysterious, and no
field theoretic formulation is known which would make it self-evident
(except in the EM case, of course.)

\section{Olive-Montonen Duality and the QCD String}

I now want to turn to the main subject of my talk, which involves the
relation between Olive-Montonen duality and QCD strings.  I will
present a variation on the Seiberg-Witten picture of confinement in
$SU(N)$ SQCD.\cite{nsewone} In this variation, the flux tubes of SQCD,
which appear in a straightforward way, carry the appropriate discrete
charges.  Previous approaches, where the charges did not work out
correctly, exhibited a spectrum that was contrary to
expectations.\cite{DS} I will explain these issues below.

\subsection{Electric Sources and Fluxes}

I begin with a review of basic facts about gauge theories.
Consider a pure gauge theory with gauge group $G$.  Suppose we have a
source --- an infinitely massive, static, electrically charged
particle --- in a representation $R$ of $G$.  If we surround the
source with a large sphere, what characterizes the flux passing
through the sphere?  If $G$ is $U(1)$, the flux measures the electric
charge directly. However, in non-abelian gauge theories the gauge
bosons carry charge.  Since there may be a number (varying over time)
of gauge bosons inside the sphere, the representation under which the
charged objects in the sphere transform is not an invariant.  But, by
definition, the gauge bosons are neutral under the discrete group
$C_G$, the center of $G$.  It follows that the charge of $R$ under the
center {\it is} a conserved quantity, and that the total flux exiting
the sphere carries a conserved quantum number under $C_G$.

For example, in $SU(N)$ the center is a $\ZN$ group
with the lovely name of ``N-ality''.  A tensor $T^{a_1a_2\cdots a_p}_{b_1b_2\cdots b_q}$ has N-ality $p-q$ mod $N$; in particular, the $k$-index
antisymmetric tensors carry N-ality $k$.

\ures Electric sources and fluxes in pure gauge theories carry a
conserved $C_G$ quantum number.  If the gauge group confines, then the
confining electric flux tubes will also carry this quantum number.

If the theory also contains light matter charged under $C_G$ but
neutral under a subgroup $C_m$ of $C_G$, then the above statements are
still true with $C_G$ replaced with $C_m$.  For example, if we take
$SU(N)$ with massless fields in the ${\bf N}$ representation, then
$C_m$ is just the identity, reflecting the fact that all sources can
be screened and all flux tubes break.  If we take $SO(10)$ with fields
in the ${\bf 10}$, then the center $\ZZ_4$ is replaced with
spinor-number $\ZZ_2$.  Sources in the ${\bf 10}$ will be screened and
have no flux tube between them, while sources in the ${\bf 16}$ or
${\bf \overline{16}}$ will be confined by a single type of flux tube.

\subsection{Magnetic Sources and Fluxes}

Before discussing the magnetic case, I review some basic topology.
The $p$-th homotopy group of a manifold $\MM$, $\pi_k(\MM)$, is the
group of maps from the $p$-sphere into $\MM$, where we identify maps as
equivalent if they are homotopic in $\MM$.  All we will need for present
purposes are the following examples.  Suppose a Lie group $G$ has rank
$r$, so that its maximal abelian subgroup is $U(1)^r$; then
\bel{pitwo}
\pi_2[G] = {\bf 1}\ \  \Rightarrow \ \
\pi_2\left[G/U(1)^r\right] 
= \pi_1[U(1)^r] =  \ZZ\times \ZZ\times \cdots \times \ZZ
\equiv [\ZZ]^r  \ .
\ee
Similarly,
\bel{pione} 
\pi_1[G] = {\bf 1} \ \  \Rightarrow \ \ \pi_1\left[G/C_G\right] =
\pi_0[C_G] = C_G\ .  
\ee

We will need to investigate both monopole solitons and string solitons
below.  The classic monopole soliton is that of 't Hooft and of
Polyakov, which arises in $SU(2)$ broken to $U(1)$; in this case the
important topological relation is $\pi_2[SU(2)/U(1)] =
\pi_1[U(1)]=\ZZ$.  This leads to a set of monopole solutions carrying
integer charge.  Note that the stability of, for example, the
charge-two monopole solution against decay to charge-one monopoles is
determined not by topology but by dynamics.  The situation is similar
for the Nielsen-Olesen magnetic flux tube of the abelian Higgs model;
here the relevant topological relation is $\pi_1[U(1)]=\ZZ$.  This
again leads to solutions with an integer charge, whose stability
against decay to minimally charged vortices is determined 
dynamically.

More generally, if we have a {\it simply connected} gauge group $G_0$
which breaks to a group $G$ at a scale $v$, there will be monopoles
carrying a quantum number in $\pi_2[G_0/G]$, of mass [radius]
proportional to $v$ [$1/v$].  Now imagine that we take $v\rarr\infty$.
In this limit the gauge group $G_0$ disappears from the system.  The
monopoles become pointlike and infinitely massive; their only
non-pointlike feature is their Dirac string, which carries a quantum
number in $\pi_1[G]$. In short, the solitonic monopoles become
fundamental Dirac monopoles in this limit. Note that since
$\pi_2[G_0/G]=\pi_1[G]$, the charges carried by the solitonic
monopoles and their Dirac monopole remnants are the same.  Since the
Dirac monopoles heavy, we may use them as magnetic sources.

Let's further suppose that the gauge group $G$ is broken completely at
some scale $v'$.  In this case no Dirac strings can exist in the
low-energy theory, and so the monopoles allowed previously have
seemingly vanished.  However, solitonic magnetic flux tubes, carrying
charges under $\pi_1[G]$, will be generated; they will have tension
[radius] of order $v'^2$ [$1/v'$].  Their $\pi_1[G]$ quantum numbers
are precisely the ones they need to confine the $\pi_1[G]$-charged
Dirac monopole sources of the high-energy theory.  Thus, when $G$ is
completely broken, the Dirac monopoles disappear because they are
confined by flux tubes.

\ures Magnetic sources and fluxes in pure gauge theories carry a
conserved $\pi_1[G]$ quantum number.  If the gauge group is completely
broken, then the confining magnetic flux tubes will also carry this
quantum number.  

\subsection{\nfour\ Supersymmetric Gauge Theory}

The next ingredient in this {\it okonomiyaki} is \nfour\ \susic\ gauge theory,
consisting of one gauge field, four Majorana fermions, and six real
scalars, all in the adjoint representation.  It is useful to combine
these using the language of \none\ \susy, in which case we have one
vector multiplet (the gauge boson $A_\mu$ and one Majorana fermion
$\lambda$) and three chiral multiplets (each with a Weyl fermion
$\psi^s$ and a complex scalar $\Phi^s$, $s=1,2,3$.)

These fields have the usual gauged kinetic terms, along with
additional interactions between the scalars and fermions.  The
scalars, in particular, have potential energy
\be
V(\Phi^s) = \sum_{a=1}^{dim \ G} |D_a^2| + \sum_{s=1}^3 |F_s|^2
\ee
 where
\be
D_a = \left(\sum_{s=1}^3 [\Phi^{s\dag},\Phi^s]\right)_a
\ee
(here $a$ is an index in the adjoint of $G$) and
\bel{Fnomass}
F_s = \epsilon_{stu}  [\Phi^t,\Phi^u] \ .
\ee
Supersymmetry requires that $\vev{V(\Phi^s)}=0$, and so all $D_a$ and
$F_s$ must vanish separately.  The solution to these requirements is
that a single linear combination $\hat \Phi$ of the $\Phi^s$ may have
non-vanishing expectation value, with the orthogonal linear
combinations vanishing.  By global rotations on the index $s$ we may
set $\hat \Phi=\Phi^3.$ By gauge rotations we may make $\Phi^3$ lie in
the Cartan subalgebra of the group; we may represent it as a diagonal
matrix
\bel{vacs}
\vev{\Phi^3} = {\rm diag}(v_1,v_2,\cdots)
\ee
which (if the $v_i$ are all distinct) breaks $G$ to $U(1)^r$.
Since $\pi_2[G/U(1)^r]= [\ZZ]^r$ [see \Eref{pitwo}] the theory
has monopoles carrying $r$ integer charges under $U(1)^r$.
Quantum mechanically, the theory has both monopoles and dyons,
carrying $r$ electric and $r$ magnetic charges $(n_e,n_m)$.    

The space of vacua written in \Eref{vacs} is not altered by quantum
mechanics.  In the generic $U(1)^r$ vacuum, each $U(1)$ has EM
duality.  In a vacuum where some of the $v_i$ are equal, the gauge
group is broken to a non-abelian subgroup $\hat G$ times a product of
$U(1)$ factors; the low-energy limit of the non-abelian part is an
interacting conformal field theory.  The $U(1)$ factors have EM
duality, while the $\hat G$ factor has its non-abelian generalization,
OM duality.\cite{omdual,gom,osborn,nsewtwo}

\subsection{Olive-Montonen Duality}

The \nfour\ theory has a set of alternate descriptions, generated
by changes of variables (whose explicit form remains a mystery)
of the form
\be
{\bf T} \ : \ \tau\rarr\tau+1 \ (\theta\rarr\theta+2\pi) \ ; \
n_e\rarr n_e+n_m,\ n_m\rarr n_m \ ; \ G\rarr G \ ;
\ee
and
\bel{Stransf}
{\bf S} \ : \ \tau\rarr-{1\over\tau} \ 
 (g\rarr {4\pi\over g} {\rm \ if \ } \theta=0) \ ; \
n_e\leftrightarrow n_m \ ; \ G\rarr \tilde G \ .
\ee
Together ${\bf S}$ and ${\bf T}$ generate the group $SL(2,\ZZ)$, which
takes $\tau\rarr (a\tau+b)/(c\tau+d)$ for integers $a,b,c,d$
satisfying $ad-bc=1$.  Note that ${\bf T}$ is nothing but a rotation
of the theta angle by $2\pi$; it does not change the gauge group or
the definition of electrically charged particles, shifting only the
electric charges of magnetically charged particles.\cite{witeff} By
contrast, ${\bf S}$ exchanges electric and magnetic charge, weak and
strong coupling (if $\theta=0$),\cite{omdual} and changes the gauge
group\cite{gom,osborn} from $G$ to its dual group $\tilde G,$ as
defined below.

The group $G$ has a root lattice $\Lambda_G$ which lies in an $r=$
rank$(G)$ dimensional vector space.  This lattice has a corresponding
dual lattice $(\Lambda_G)^*$.  It is a theorem that there exists a Lie
group whose root lattice $\Lambda_{\tilde G}$ equals
$(\Lambda_G)^*$.\cite{gom} Here are some examples:
\be 
\matrix{{SU(N)}\leftrightarrow SU(N)/\ZN \ ; & \ 
{SO(2N+1)}\leftrightarrow USp(2N) \ ; \cr 
{SO(2N)}\leftrightarrow SO(2N)   \ ; & \ 
{Spin(2N)} \leftrightarrow  SO(2N)/\ZZ_2 \ .}
\ee
Notice that this set of relationships depends on the global
structure of the group, not just its Lie algebra; $SO(3)$ (which
does not have spin-$1/2$ representations) is dual to $USp(2)\approx SU(2)$
(which does have spin-$1/2$ representations.)  These details are essential
in that they affect the topology of the group, on which OM duality depends.

In particular, there are two topological relations which are of great
importance to OM duality.  The first is relevant in the generic
vacuum, in which $G$ is broken to $U(1)^r$.  The electric charges
under $U(1)^r$ of the massive electrically charged particles (spin
$0,\half,1$) lie on the lattice $\Lambda_G$. The massive magnetic
monopoles ({\it also} of spin $0,\half,1$) have magnetic charges under
$U(1)^r$ which lie on the dual lattice
$(\Lambda_G)^*$.\cite{gom,osborn} Clearly, for the $S$ transformation,
which exchanges the electrically and magnetically charged fields and
the groups $G$ and $\tilde G$, to be consistent, it is essential that
$\Lambda_{\tilde G} = (\Lambda_G)^*$ --- which, fortunately, is true.

The second topological relation is the one we will use below.  We have
seen that the allowed electric and magnetic sources for a gauge theory
with adjoint matter (such as \nfour) are characterized by quantum
numbers in $C_G$ and $\pi_1(G)$ respectively.  Consistency of the
${\bf S}$ transformation would not be possible were these two groups
not exchanged under its action.  Fortunately, it is a theorem of group
theory that\cite{gom}
\bel{piGCG}
\pi_1(G) = C_{\tilde G} \ ; \ \pi_1(\tilde G)= C_G \ .
\ee
For example, $\pi_1[SU(N)] = C_{SU(N)/\ZN} = {\bf 1}$ while
$C_{SU(N)} = \pi_1[SU(N)/\ZN] = \ZN$.

\ures As a consequence of \Eref{piGCG} and the results of sections 4.1
and 4.2, the allowed magnetic sources of $G$ are the same as the
allowed electric sources for $\tilde G$, and vice versa.

\subsection{Breaking \nfour\ to \none}

 Now, we want to break \nfour\ \susy\ to \none.  Pure \none\ SQCD,
like pure non-\susic\ QCD, is a confining theory, and should contain
confining flux tubes.  The addition of massive matter in the adjoint
representation does not change this; heavy particles would only obstruct
confinement by breaking flux tubes, which adjoint matter cannot do.
We therefore expect that broken \nfour\ gauge theory, which is \none\
SQCD plus three massive chiral fields in the adjoint representation,
should be in the same universality class as pure SQCD: both should
confine, and both should have flux tubes carrying a $C_G$ quantum
number, as discussed in section 4.1.

We may break the \nfour\ symmetry by adding masses $m_s$ for
the fields $\Phi^s$; the $F_s$ functions of \eref{Fnomass} become
\bel{Fwmass}
F_s = \epsilon_{stu} [\Phi^t,\Phi^u] + m_s\Phi^s \ ,
\ee
so that $F_s=0$ implies
$\epsilon_{stu}[\Phi^t,\Phi^u]=-m_s\Phi^s$.\cite{cvew} Up to
normalization these are the commutation relations for an $SU(2)$
algebra, which I will call $SU(2)_{aux}$.  If we take $m_1=m_2=m$ and
$m_3=\mu$ we obtain
\bel{Phivevs}
\Phi^1 = -i\sqrt{\mu m} J_x \ ;
\Phi^2 = -i\sqrt{\mu m} J_y \ ;
\Phi^3 = -i m J_z \ ,
\ee
where $J_x,J_y,J_z$ are matrices satisfying $[J_x,J_y]=iJ_z$, {\it
etc.}  Each possible choice for the $J$'s gives a separate, isolated
vacuum.\cite{cvew}

How does this work, explicitly, in $SU(N)$?  We can write the $\Phi^s$
as $N\times N$ traceless matrices, so the $J_i$ should be an
$N$-dimensional (generally reducible and possibly trivial)
representation of $SU(2)_{aux}$.\cite{cvew,rdew}  The trivial choice
corresponds to $J_i=0$; clearly if $\Phi^s=0$ the $SU(2)_{aux}$
commutation relations are satisfied.  We will call the corresponding
vacuum the ``unbroken'' vacuum, since the $SU(N)$ gauge group is
preserved.  Another natural choice is to take the $J_i$ in the
irreducible spin-${N-1\over2}$ representation of $SU(2)_{aux}$.  In
this case $SU(N)$ is completely broken; we will call this the ``Higgs
vacuum''.  We may also choose the $J_i$ in a reducible representation
\be
J_i= \left[\matrix{\sigma_i & | &0\cr  --&-|-&--\cr 0& | &0 }\right] \ ;
\ee
here the $\sigma_i$ are the Pauli matrices.
In this case $SU(N)$ is partly broken.  There are many vacua like this
last one, but they will play no role in the physics below; we will
only need the unbroken vacuum and the Higgs vacuum.

\ures The classical analysis of this \none\ \susic\ $SU(N)$ gauge
theory with massive adjoint fields shows that it has isolated \susic\
vacua scattered about, with the unbroken (U) vacuum at the origin of
field space and the Higgs vacuum (H) at large $\Phi^s$ expectation
values [of order $m,\sqrt{m\mu}$, see \Eref{Phivevs}.]\cite{cvew,rdew}

\subsection{OM Duality and the Yang-Mills String}

The above picture is modified by quantum mechanics.  In each vacuum,
strong dynamics causes confinement to occur in the unbroken
non-abelian subgroup, modifying the low-energy dynamics and generally
increasing the number of discrete vacua.  In the H vacuum, the gauge
group is completely broken and no non-trivial low-energy dynamics
takes place; it remains a single vacuum.  The U vacuum, by contrast,
splits into $N$ vacua --- the well-known $N$ vacua of
SQCD\cite{Nvacua} --- which I will call ${\rm D}_0,{\rm
D}_1,\cdots,{\rm D}_{N-1}$. In the ${\rm D}_k$ vacuum,
confinement occurs by condensation of dyons of magnetic charge 1 and
electric charge $k$.\cite{nsewone,rdew,DS,sun}  Since these vacua are
related\cite{Nvacua} by rotations of $\theta$ by multiples of $2\pi$,
I will focus on just one of them.  It is convenient to study the ${\rm
D}_0$ vacuum (which I now rename the M vacuum) in which electric
charge is confined by magnetic monopole condensation.

Now, what is the action of OM duality on this arrangement?  The vacua
are physical states, and cannot be altered by a mere change of
variables; however, the {\it description} of each vacuum will change.
Specifically, when $\theta=0$, the ${\bf S}$ transformation, which
inverts the coupling constant and exchanges electric and magnetic
charge, exchanges the H vacuum of $SU(N)$ for the M vacuum of
$SU(N)/\ZN$ and vice versa.\cite{rdew}  To say it another way,
the confining M vacuum of $SU(N)$ can be equally described as the H
vacuum of $SU(N)/\ZN$, in which the monopoles of the $SU(N)$
description break the dual $SU(N)/\ZN$ gauge group.  This is the
Generalized Dual Meissner effect, in which both the electric and
magnetic gauge groups are non-abelian.

\ures OM duality exchanges the H and M vacua of $SU(N)$ broken \nfour\
with the M and H vacua of $SU(N)/\ZN$ broken \nfour.  Confinement in
the M vacuum of $SU(N)$ is described as the breaking of the
$SU(N)/\ZN$ gauge group in $SU(N)/\ZN$.\cite{rdew}

The existence of the Yang-Mills string now follows directly from
topology.  As we discussed in section 4.2, the complete breaking of a
group $G$ leads to solitonic strings carrying magnetic flux with a
quantum number in $\pi_1(G)$.  In this case, the breaking of
$SU(N)/\ZN$ in its H vacuum (the confining M vacuum of $SU(N)$) gives
rise to strings with a $\ZN$ quantum number.  But magnetic flux tubes
of $SU(N)/\ZN$ are, by OM duality, electric flux tubes of $SU(N)$ ---
and so the confining strings of the $SU(N)$ theory's M vacuum, the
confining theory which is in the same universality class as $SU(N)$
SQCD, carry a $\ZN$ quantum number.  This is in accord with the
considerations of section 4.1.  The relation \Eref{piGCG} is
responsible for this agreement of the $\ZN$ charges, and presumably
assures a similar agreement for all groups.

\ures OM duality gives a picture for confinement in $SU(N)$ SQCD ---
it occurs via {\it non-abelian} dual monopole condensation, and leads
to confining strings with a $\ZN$ quantum number.

A cautionary remark is in order.  The description of confinement via
dual monopole condensation is not fully reliable, as it is only
appropriate if the $SU(N)/\ZN$ theory is weakly coupled.  In fact, we
want the $SU(N)$ theory to be weakly coupled in the ultraviolet, in
analogy with QCD.  The ${\bf S}$ transformation implies that the
$SU(N)/\ZN$ description should be {\it strongly} coupled in the
ultraviolet.  However, the existence of a soliton carrying a stable
topological charge is more reliable, especially since there are no
other objects carrying that charge into which these string solitons
can decay.  Having constructed the solitonic strings semiclassically
in some regime, we expect that they survive into other regimes in
which semiclassical analysis would fail. (A gap in the argument: could
the strings grow large and have zero string tension in the SQCD
limit?)  In short, while the condensing monopole description
appropriate to broken \nfour\ SQCD may not be valid for pure \none\
SQCD, it does demonstrate the presence of confining strings in the
latter.  Whether anything quantitative can be said about these strings
is another matter, to be addressed below.

Should we expect this picture to survive to the non-\susic\ case?
Take the theory with \nfour\ \susy\ broken to \none, and futher break
\none\ \susy\ by adding an $SU(N)$ gaugino mass $m_\lambda\ll m,\mu$.
We cannot be sure of the effect on the dual $SU(N)/\ZN$ theory;
duality does not tell us enough. However, we know that the theory has
a gap, so this \susy-breaking can only change some properties of the
massive fields, without altering the fact that $SU(N)/\ZN$ is
completely broken.  The strings, whose existence depends only on this
breaking, thus survive for small $m_\lambda$.  To reach pure QCD
requires taking $m,\mu,m_\lambda$ all to infinity.  It seems probable,
given what we know of QCD physics, that the strings undergo no
transition as these masses are varied.  In particular, there is
unlikely to be any phase transition for the strings between pure SQCD
and pure QCD; this conjecture can and should be tested on the lattice.

\ures If the strings of SQCD and of QCD are continously related,
without a transition as a function of the gaugino mass, then the
arguments given above for SQCD extend to QCD, establishing a direct
link between OM duality of \nfour\ gauge theory and the confining
$\ZN$-strings of pure QCD.

\subsection{Confinement According to Seiberg and Witten}

How does this picture of confinement differ from that of Seiberg and
Witten?  Where and why might it be preferable?

Seiberg and Witten studied pure \ntwo\ \susic\ $SU(2)$ gauge
theory.\cite{nsewone} They showed that the infrared quantum mechanical
theory could be understood as a $U(1)$ theory coupled to a magnetic
monopole, and that, when \ntwo\ \susy\ is broken to \none, the
monopole condenses, confining the $SU(2)$ degrees of freedom.  The
picture generalizes\cite{sun} to $SU(N)$, where the infrared physics involves
$U(1)^{N-1}$ coupled to $N-1$ monopoles, whose condensation drives
confinement.  This was studied in detail by Douglas and Shenker.\cite{DS}

This physics is contained in a particular regime of the broken \nfour\
\susy\ gauge theory discussed above.  If we take $\mu=0$, then the
theory is \ntwo\ \susic, and, as seen from \Eref{Phivevs}, the H
vacuum of $SU(N)/\ZN$ has its gauge group broken only to $U(1)^{N-1}$.
Now take $m$ exponentially large and the coupling at the scale $m$
small, so that the strong coupling scale $\Lambda$ of the low-energy
theory is finite, and let $\mu$ be non-zero but small compared to
$\Lambda$. The low-energy confining vacua of the $SU(N)$ theory will
then be the vacua studied by Douglas and Shenker.  The magnetic
description of the theory (using OM duality) will have the gauge group
$SU(N)/\ZN$ broken at a high scale to $U(1)^{N-1}$, which in turn is
broken completely at a low scale; see \Eref{Phivevs}.  The second step
in this breaking leads to confinement of $SU(N)$ fields.  If we take
$m$ to infinity, then the $SU(N)/\ZN$ gauge group disappears from the
theory.  The magnetic theory is merely $U(1)^{N-1}$ broken to nothing,
as in Douglas and Shenker.\cite{sun,DS} (Note that I am cheating a bit
here, as the abelian theory requires a cutoff; I'll fix this below.)

Since the magnetic theory is $U(1)^{N-1}$, dual to the maximal abelian
subgroup of $SU(N)$, the pure \ntwo\ theory exhibits a dynamical form
of abelian projection.\cite{DS}  The monopoles, whose condensation drives
confinement when \ntwo\ \susy\ is weakly broken, are purely abelian.
Given the number of talks at this conference on abelian projection as
a mechanism for explaining confinement, why should this disturb us?

The problem lies with the quantum numbers of the strings.  The
$U(1)^{N-1}$ magnetic theory consists of $N-1$ copies of the abelian
Higgs model, each of which has a \NO\ solitonic flux tube.\cite{DS}
These strings carry quantum numbers in the group
$\pi_1[U(1)^{N-1}]=[\ZZ]^{N-1}$, {\it not} in $\ZN$!  That is, each of
the $N-1$ Nielsen-Olesen strings carries its own conserved integer
charge. These strings cannot lead to a good model for the dynamics of
$SU(N)$ SQCD or QCD, which on general grounds must have $\ZN$-carrying
strings.

\begin{figure}
\centering
\epsfxsize=5.5in
\hspace*{0in}
\vspace*{.2in}
\epsffile{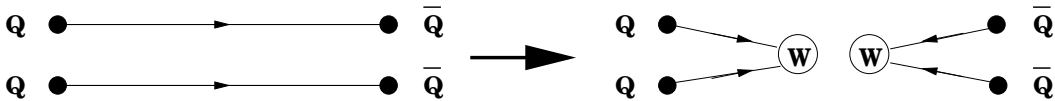}
\caption{A pair of parallel $SU(2)$ strings can break via $W$ boson 
production.}
\label{fig:stringbreak}
\end{figure}

Is this problem serious?  At first glance, the little cheat that I
made just a moment ago rescues the abelian projection.  The magnetic
$U(1)^{N-1}$ theory has a cutoff at the scale $\Lambda$, where its
coupling becomes large.  At that scale, there are massive
electrically-charged gauge bosons of $SU(N)$.  Pair production of
these particles cause certain configurations of parallel strings,
which would naively be stable according to the reasoning of the
previous paragraph, to break.  The charges of these gauge particles
are precisely such that they reduce the conserved symmetry from
$[\ZZ]^{N-1}$ to $\ZN$.  For example, consider the case of $SU(2)$, as
shown in figure \ref{fig:stringbreak}.  An isospin-$\half$ quark and a
corresponding antiquark will be joined by a Nielsen-Olesen string.  This
string is stable.  However, two such quark-antiquark pairs, with
parallel strings, are unstable to reconfiguring their strings via $W$
boson production.  This reflects the claim above that $W$ production
reduces the symmetry under which the $SU(2)$ strings transform from
$\ZZ$ to $\ZZ_2$.

\

\

Topologically speaking, all seems well --- the charges seem to be the
expected ones --- but the dynamics of the theory still poses a serious
problem.  Although $[\ZZ]^{N-1}$ is not exactly conserved, it is {\it
approximately} conserved.  To see this in the $SU(2)$ example, note
that the $W$ pair-production requires an energy of order $\Lambda$.
The string tension (its energy per unit length) is $T =
R_{conf}^{-2}\approx \mu\Lambda$ in this theory; here $R_{conf}$ is
the confinement length.  In order for the string to have enough energy
to break, it should have length $L$ such that $TL\sim\Lambda$.  This
implies
\bel{LvsR}
L \geq {1\over\mu} \gg {1\over \sqrt{\mu\Lambda}} = R_{conf} \ ,
\ee
so only enormously long strings can break.  Furthermore, since the
strings' energy density is very low, it takes a large fluctuation to
generate $W$ boson pairs.  This in turn means that the rate for the
transition in figure \ref{fig:stringbreak} is very slow.  In short,
the string pair shown in the figure is {\it metastable} for
$\mu\ll\Lambda$.  The $\ZZ$ symmetry is still approximately conserved
by the dynamics.  Similar arguments apply for $SU(N)$.

\ures The strings of weakly broken \ntwo\ $SU(N)$ SQCD carry an approximately
 conserved $[\ZZ]^{N-1}$ symmetry, which contains the expected $\ZN$
 as an exactly conserved subgroup.  This leads to metastable string
configurations not expected in \none\ SQCD and in QCD.

The physical consequences of this approximately conserved symmetry are
potentially dramatic.  In SQCD, a pair of parallel strings which carry
fluxes of charge $k$ and $p$ under $\ZN$ should undergo a rapid
transition to a string carrying charge $k+p$ mod $N$.  (In QCD, this
implies that the string between a quark and an antiquark is the same
as the string between a quark and a diquark.)  But this transition is
inhibited in the broken \ntwo\ gauge theories for small $\mu$. This
implies that numerous, distinct, metastable configurations of strings
may connect a quark in the ${\bf N}$ representation to a corresponding
antiquark.  For example, for $k=1,\cdots N/2$, a pair of parallel
strings, with charges $k$ and $N-k+1$ respectively, carry total charge
$1$; they therefore may, as a pair, join a quark and antiquark.
(Since a string with charge $N$ is no string at all, the $k=1$ case is
the expected one.)  These $N/2$ metastable configurations have
different energies per unit length, and in principle can give
physically distinct quark-antiquark meson Regge trajectories.

Indeed, in \ntwo\ SQCD, the dynamics of the theory breaks the Weyl
group, so the $N$ colors of quark are inequivalent.  As shown by
Douglas and Shenker, each color of quark prefers a {\it different}
choice of string pairs.\cite{DS}  This leads to their most surprising
conclusion.

\ures In weakly broken \ntwo\ \susic\ $SU(N)$ gauge theory, the
quark-antiquark mesons exhibit $N/2$ Regge trajectories,\cite{DS}
instead of one as expected in \none\ SQCD and in QCD.

What happens to these extra trajectories as $\mu\rarr\infty$ and the
theory approaches pure \none\ SQCD?  As can be seen from \Eref{LvsR},
the obstructions to $W$ pair production go away as $\mu\rarr\Lambda$.
(Note the formulas which lead to \eref{LvsR} receive corrections at
order $\mu/\Lambda$.)  The extra Regge trajectories become highly
unstable and disappear from the spectrum.\cite{DS} There is no sign of
conflict with the usual SQCD expectations of a single Regge trajectory
and of strings with a $\ZN$ symmetry.  However, the $U(1)^{N-1}$
magnetic theory which we used to describe the weakly broken \ntwo\
theory becomes strongly coupled in this limit, and so one cannot study
this picture quantitatively.

In summary, although broken \ntwo\ \susic\ gauge theory can be used to
show that \none\ SQCD is a confining theory,\cite{nsewone,sun,DS} it
is not a good model for the hadrons of \none\ SQCD.  This is a direct
consequence of the dynamical abelian projection, which leads to an
abelian dual description.  The condensation of its abelian monopoles
leads to confinement by Nielsen-Olesen strings, which carry
(approximately) conserved integer charges that (S)QCD strings do not
possess.  These charges alter the dynamics of bound states, leading to
a spectrum and to hadron-hadron interactions very different from those
expected in SQCD and found in QCD.  By contrast, these problems are
avoided in the broken \nfour\ description of confinement given in
sections 4.5-4.6.

{\underline {Moral:}} The use of abelian projection, and the
construction of a dual abelian gauge theory, has inherent difficulties
in explaining the dynamics of QCD strings.  One must therefore
use abelian projection with caution.  It may provide good answers for
a limited set of questions, but for other questions it may fail badly.

\subsection{String tensions in $SU(N)$}

The discussion to this point has been entirely
qualitative.  Are any quantitative predictions possible?

A very useful theoretical quantity to study is the ratio of tensions
of strings carrying different charge under $\ZN$.  A confining string
of $SU(N)$ SQCD or QCD with quantum number $k$ under $\ZN$ has a
tension $T_k$ which depends on $k,N$ and the strong coupling scale
$\Lambda$.  On dimensional grounds $T_k = \Lambda^2 f(k,N)$.  While no
analytic technique is likely to allow computation of $T_1$, it is
possible that $T_k/T_1$ can be predicted with some degree of accuracy.
Note that charge conjugation implies $T_k = T_{N-k}$, so $SU(2)$ and
$SU(3)$ have only one string tension.  We must consider $SU(4)$ and
higher for this to be non-trivial.

While the ratio of tensions cannot be computed in continuum SQCD or
QCD, it has been calculated in theories which are believed to be in
the same universality class.  These theories often have multiple mass
scales $\mu_i$ and thus in principle it may be that $T_k = \Lambda^2
h(\mu_i/\Lambda,k,N)$.  However, in all cases studied so far, it has
been found that $T_k = g(\mu_i,\Lambda) f(k,N)$, where $f$ is
dimensionless, and $g$ is a dimensionful function which is independent
of $k$.  Note $g$ cancels out in ratios of tensions.  Thus our attention may
focus on the dimensionless function $f(k,N)$ as a quantity which may
be compared from theory to theory.

Some previous calculations include the well-known
strong-coupling expansion of QCD, which to leading order
gives
\bel{fsc}
f_{sc}(k,N) \propto k(N-k) \ ,
\ee
and the results of Douglas and Shenker for weakly
broken \ntwo\ gauge theory\cite{DS}
\bel{fds}
f_{DS}(k,N) \propto \sin{\pi k\over N} \ .
\ee
The considerations of sections 4.5-4.6 suggest another calculation: if
the string solitons in broken \nfour\ $SU(N)$ gauge theory were
computed, the ratios of their tensions would be of considerable
interest.  At the time of writing these soliton solutions have not
appeared in the literature.

There is one other technique by which string soliton tensions may be
computed, using M theory!  M theory is eleven-dimensional supergravity
coupled to two-dimensional membranes.  The theory also has
five-dimensional solitons, called ``five-branes'', whose worldvolume
is six-dimensional.  The theory on the worldvolume of the five-brane
is poorly understood but is known to be a 5+1 dimensional conformal
field theory.

\begin{figure}
\centering
\epsfxsize=3in
\hspace*{0in}
\epsffile{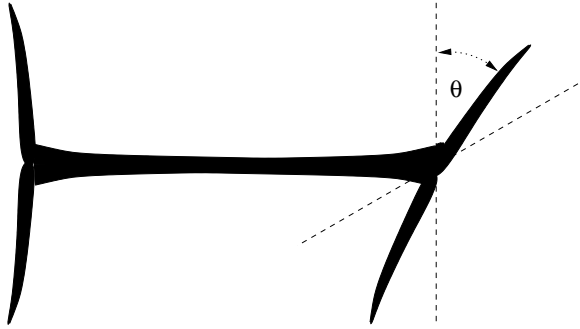}
\caption{The M theory five-brane, with four of its dimensions filling
spacetime (not shown) and the other two dimensions embedded in a
six-dimensional space, of which three dimensions are shown.  The
\ntwo\ \susy-breaking parameter $\mu$ is proportional\cite{barbon} to
$\tan \theta$; thus $\theta=0$ has \ntwo\ \susy\ while $\theta=\pi/2$
gives \none\ SQCD.}
\label{fig:Mbrane}
\end{figure}

As shown by Witten,\cite{witMa,witMb} following work of Elitzur,
Giveon and Kutasov,\cite{EGK} one may consider M theory on
$\MM^{10}\times S^1$, where $\MM^{10}$ is ten-dimensional Minkowski
space and $S^1$ is a circle of radius $R_0$.  One can construct
five-branes with rather strange shapes (figure \ref{fig:Mbrane}) whose
worldvolume theory contains, at low-energy, a sector with the same
massless fields and interactions as \ntwo\ $SU(N)$ SQCD, or broken
\ntwo\ $SU(N)$ SQCD, or pure \none\ SQCD.  (These theories also
contain an infinite tower of massive particles, all neutral under
$\ZN$; thus they are potentially in the same universality class as,
but should not be confused with, the gauge theories we are interested
in.\cite{witMb}) Witten showed membranes can bind to these
five-branes, making objects that carry a $\ZN$ charge and look in 3+1
dimensions like strings.\cite{witMb} It was then shown\cite{hsz} that
these strings indeed confine quarks and reproduce the results of
Douglas and Shenker in the appropriate limit.  However, the string
tension ratios can be computed (naively) even when the \ntwo\ breaking
parameter $\mu$ is large. One finds that the tensions are
given by\cite{hsz}
\bel{TM}
T_k = g(\mu,\Lambda,R_0) f_{DS}(k,N)
\ee
where $g$ is a complicated dimensionful function which cancels out of
tension ratios, and $f_{DS}$ is as in \Eref{fds}.

Thus, M theory suggests that the tensions ratios satisfy \eref{fds}
for large as well as small breaking of \ntwo\ \susy.
One must be careful with this result, however. First, as 
mentioned above, this result applies for the M theory version
of SQCD, which has extra massive states that normal SQCD
does not have.\cite{witMb}  Second, no non-renormalization theorem
protects the result \eref{TM} when $\mu$ is large.
The overall coefficient function $g$ is certainly 
renormalized.  The question is whether $f$ is
strongly renormalized or not. Only a lattice computation
will resolve this issue.

What about non-\susic\ QCD?  A very naive computation in M theory
suggests that $T_k\propto f_{DS}(k,N)$ even in this case!  However,
corrections from various sources are expected, and the accuracy of
this prediction is in doubt.\cite{hsz,esbrane}

In summary, we so far have results on confining strings from the
strong coupling expansion on the lattice (which only exists for
non-\susic\ theories), from broken \ntwo\ \susic\ gauge theory in the
continuum, and from the M theory versions of QCD and SQCD.  There are
a couple of interesting observations worth making.

First, in all of the calculable limits, $T_k/T_1<k$ for all $k$; thus
a string with charge $k$ is stable against decay to $k$ strings of
charge one.  Consequently, we expect that these flux tubes attract one
another, and that therefore QCD and SQCD correspond to type I dual
superconductors.  I believe this result is robust and will be
confirmed numerically.

Second, the functions $f_{sc}$ and $f_{DS}$ have different large $N$
behavior.  They agree at leading order, but while the first correction
is at order $1/N$ for $f_{sc}$ (as we would expect for an $SU(N)$
theory), the first correction for $f_{DS}$ is order $1/N^2$!  The fact
that the $1/N$ correction in $f_{DS}$ {\it vanishes} is surprising,
and the physics that lies behind this feature has not been explained.

\section{A Similar Case}

As an aside, I would like to mention the earlier work in which
the first concrete example of the non-abelian generalized dual Meissner
effect was presented.\cite{spinmono}

Consider \none\ \susic\ $SO(N)$ gauge theory with $N_f$ matter fields
in the ${\bf N}$ representation.  In this case, electric sources in
the spinor representation of $SO(N)$ can be introduced; they carry a
$\ZZ_2$ quantum number --- ``spinor number'' --- which obviously
cannot be screened by adjoints or vectors of $SO(N)$.  These sources
can be used to test the phase of the $SO(N)$ theory.

On the other hand, certain facts are known about these $SO(N)$ gauge
theories. In particular, depending on $N$ and $N_f$,
the theory may be in any of the phases discussed in section 2. We also
know the $SO(N)$ duality relations discussed in section 2.  The
strong-coupling physics of the electric $SO(N)$ theory is often best
understood using its magnetic description, particularly in the free
magnetic and confining phases.  But how are the spinor sources, which
we need to test the electric theory, mapped into the dual theory?  The
original work on $SO(N)$ duality\cite{NAD,kinstwo} did not answer this
question.

The answer comes from duality itself.  Pouliot and I found a dual
description to $SO(8)$ with $N_f$ vectors and one
spinor.\cite{ppmsone} It turns out that giving a mass to the spinor
allows a monopole soliton to form in the dual theory.  This monopole
carries a $\ZZ_2$ charge, the same as the unscreened discrete charge
of the spinor.  Numerous checks confirm that the monopole {\it is} the
image under duality of the massive spinor. By making the spinor
particles very heavy, we can effectively convert them to sources.  At
the same time, the monopole becomes a $\ZZ_2$ Dirac monopole which can
be used as a magnetic source in the dual theory.

We thus learn that a Wilson loop in the spinor representation is dual
to a $\ZZ_2$ valued 't Hooft loop.  We can now use this fact to study
the phases of $SO(N)$.  Two main results are found.\cite{spinmono}

1) There has been debate as to whether electric charge is confined in
the free magnetic phase.  This phase has massless two-particle bound
states, along with non-polynomial composite gauge bosons and matter.
Intriligator and Seiberg proposed\cite{kinsrev} that electric charges,
rather than being confined, should have a potential $\ln r/r$ at large
distance; this function is the potential energy between two monopoles
in a theory with a weak running coupling.  The mapping of the electric
spinors to massive monopoles in the infrared-free dual theory allows
their suggestion to be confirmed.

2) When the number $N_f$ of vectors is reduced sufficiently that the
free magnetic phase passes to the confining phase, the non-abelian
dual group is completely broken.  In the dual description, $\ZZ_2$
magnetic sources are confined by a string soliton carrying a $\ZZ_2$
charge.  It follows that spinor sources in the electric theory are
confined by $\ZZ_2$ electric flux tubes.  This is an example of the
generalized dual Meissner effect: confinement by a string soliton in a
non-abelian dual description, following condensation of non-abelian
monopoles.

\section{Outlook for QCD}

I'd like to finish by summarizing the questions that I've raised in
this talk, and by focussing attention on three of them that I
believe can be studied using lattice gauge theory.

\subsection{Questions from the Superworld}

First and foremost, what {\it is} duality?  We still have no explicit
understanding of non-trivial duality transformations in three or four
dimensions, and not nearly enough even in two dimensions.  Do we need
a reformulation of field theory itself?  What does duality in string
theory teach us?

  What is the phase structure of QCD as a function of gauge group,
matter content, and interactions?  Does QCD have duality similar to
SQCD?  Is there a free magnetic phase? Considerable thought needs to be put
into the question of how to address these issues effectively; see
the discussion below.

Does Olive-Montonen duality imply confinement in non-supersymmetric
Yang-Mills theory by $\ZN$-carrying electric flux tubes?  We have seen
in this talk that it does so for \none\ SQCD.  The additional tests
needed to complete the story for non-\susic\ QCD are discussed
below.

Are condensing {\it non-abelian} monopole-like operators responsible
for confinement in QCD?  Many speakers at this conference have been
seeking or discussing {\it abelian} monopole-like operators using
abelian projection.  Should a different approach be taken?

What are the $\ZN$ string tensions in pure $SU(N)$ QCD?  Do the ratios
of tensions fit any known formulas, such as those of \Eref{fsc} or
\Eref{fds}?  These questions can and should be addressed numerically;
see the discussion below.

Finally, we have seen repeatedly in this talk that massive matter,
when added to a theory, can make aspects of its physics easier to
understand.  What matter (or additional interactions) might we add to
non-supersymmetric QCD to make some of these questions more accessible
either analytically or numerically?

This is certainly not a complete list of questions --- I have
not mentioned the recent interest in axionic domain walls, for
example --- but for finiteness I'll stop here.

\subsection{Proposals for the lattice}

There are  several projects that I hope will be undertaken by the
lattice community in relation to what I have said in this talk.

1) Tension ratios: to my knowledge, no one has ever computed the ratio
of string tensions in $SU(4)$.  It would be helpful to know the ratio
of the tension $T_1$ of the string between two sources in the ${\bf
4}$ and ${\bf \bar 4}$ representation and the tension $T_2$ of the
string which confines sources in the ${\bf 6}$ representation.  One
needs to be careful to make sure the sources are far enough apart that
the ratio is approaching the asymptotic value which it attains for
infinitely long strings.  (For example, at distances of order
$\Lambda_{QCD}^{-1}$, the string connecting sources in the ${\bf 10}$
and ${\bf \overline{10}}$ will differ from the string connecting two
sources in the ${\bf 6}$; only at long distances will their tensions
agree.)  As a first step, it will be interesting to test the
prediction that $SU(4)$ QCD is a type I dual superconductor by
checking whether $T_2<2T_1$. To go further, one needs high accuracy to
distinguish the predictions of, for example, the functions $f_{sc}$
and $f_{DS}$, which only differ by ten percent for this case.  Ideally
one would even do this for $SU(5)$ and $SU(6)$, but obviously this is
a much longer-term goal.

2) The transition from SQCD to QCD: I hope that it will soon be
realistic to simulate $SU(2)$ or $SU(3)$ QCD with a Majorana spinor,
of mass $m_\lambda$, in the adjoint representation.\cite{latSQCD}
Tuning the mass to zero to make the theory \none\ SQCD may be
difficult, but many properties of the theory, including its strings,
should not be sensitive to $m_\lambda$ as long as it is much smaller
than the confinement scale $\Lambda$.  It would be very interesting to
map out the behavior of the theory, including the strings and their
tensions $T_k$, as a function of $m_\lambda/\Lambda$.  As I mentioned
earlier, the absence of a transition in the properties of the strings
would establish the linkage I have proposed between Olive-Montonen
duality and the $\ZN$ strings of $SU(N)$ QCD.

3) Phase structure of QCD: A more difficult goal is that discussed in
section 2, to map out the low-energy phases of non-supersymmetric QCD
as a function of gauge group, matter content, and interactions
(including non-renormalizable ones.)  The results described by Kanaya
in this conference are a step in that direction, but there is still far to
go. Since it is not easy for a standard lattice calculation to deal
with theories which have slow-running coupling constants, and since
such behavior is a common feature of SQCD theories in the most
interesting phases, I suspect that more sophisticated analytic and
numerical techniques are needed than are presently available.  I hope
that some of you will be motivated to address this problem.  Since it
is possible that understanding these issues in QCD will be essential
for explaining aspects of the real world, it seems to me that this
proposal is much more than an academic exercise.

Again, this is far from a complete list, but I'm sure this is 
plenty for lattice practioners to chew on. I hope there will
soon come a time when lattice studies provide us with important
quantitative and qualitative information about the behavior of
these still poorly understood theories.

\section*{Acknowledgements}
I would like to thank the organizers, especially Professor Suzuki, for
inviting me to this conference, as well as the many participants with
whom I had stimulating discussions.  I am grateful to many colleagues
at the Institute for Advanced Study for conversations, including
M. Alford, D. Kabat, A. Hanany, J. March-Russell, N. Seiberg,
F. Wilczek, and E. Witten.  This work was supported by the National
Science Foundation under Grant PHY-9513835 and by the WM Keck
Foundation.



\end{document}